\documentclass[sigconf,balance=false]{acmart}
\usepackage{popets}
\usepackage{soul}

\setcopyright{popets}
\copyrightyear{YYYY}

\acmYear{YYYY}
\acmVolume{YYYY}
\acmNumber{X}
\acmDOI{XXXXXXX.XXXXXXX}
\acmISBN{}
\acmConference{Proceedings on Privacy Enhancing Technologies}
\usepackage[symbol]{footmisc}

\newcommand{\mypara}[1]{\vspace{1.5pt}\noindent{\bf {#1}}}
\newcommand\blfootnote[1]{%
  \begingroup
  \renewcommand\thefootnote{}\footnote{#1}%
  \addtocounter{footnote}{-1}%
  \endgroup
}

\begin{document}

\title{Extended Abstract: Shaperd: Easily Adoptable Real-Time Traffic Shaper for Fully Encrypted Protocols}


\author{Sarah Wilson}
\authornote{Authors contributed equally to this work.}
\orcid{}
\affiliation{%
  \institution{University of Waterloo}
  \city{}
  \state{}
  \country{}}
\email{s5wilson@uwaterloo.ca}

\author{Stella Tian}
\authornotemark[1]
\affiliation{%
  \institution{University of Waterloo}
  \city{}
  \country{}}
\email{jy7tian@uwaterloo.ca}

\author{Sina Kamali}
\affiliation{%
  \institution{University of Waterloo}
  \city{}
  \country{}
}
\email{sinakamali@uwaterloo.ca}

\renewcommand{\shortauthors}{Wilson et al.}

\begin{abstract}
  Fully encrypted protocol-based tools (FEPs) are tools commonly used to circumvent censorship in restrictive regions, valued for their performance and security. However, in recent years, censors have been able to block them using an array of attacks based on passive traffic analysis and active probing. We propose Shaperd, an easily adoptable and real-time traffic shaper designed specifically to aid FEPs become more resilient to detection. Shaperd operates directly on packet contents in real time, using a novel constraint system to allow its users to generate traffic flows with any desired features. Our preliminary results reveal Shaperd introduces minimal overhead to the underlying system's throughput.
\end{abstract}


\maketitle
\blfootnote{Extended abstract presented at Free and Open Communications on the Internet (FOCI 2025), held in conjunction with the 25th Privacy Enhancing Technologies Symposium (PETS 2025).}

\section{Introduction}

Over time, more countries are resorting to internet censorship in times of political unrest, protests, or elections \cite{hoang2021great, aryan2013internet}. This restriction on the information flow results in the censor's complete control over the public's perception of ongoing events, which could strongly impact the outcome of the situation.

To thwart internet censorship attempts, users have to resort to censorship circumvention (CC) systems. Many CC systems have been proposed over the years, with various degrees of practicality, effectiveness, and performance \cite{tschantz2016sok,khattak2016sok}. These tools range from solutions such as decoy routing \cite{bocovich2016slitheen}, which are potentially effective and resilient, but are hard to deploy, to data embedding solutions such as multimedia protocol tunneling \cite{mohajeri2012skypemorph,figueira2022stegozoa,jia2023voiceover}, which are considered secure, but lack the required performance for daily activities.

One of the most used and widely adopted CC systems are tools based on fully encrypted protocols (FEPs) \cite{fenske2024bytes,ji2022security}. FEPs function by encrypting the entirety of the traffic payload, making the traffic appear as random bytes. This approach allows FEPs to effectively hide traffic metadata, such as the underlying protocol and data length, which enables them to avoid censoring tactics that rely on protocol or domain blocklisting \cite{spring_2015} as the censor cannot recognize them. Doing so effectively makes the data sent by FEPs "look random" \cite{dixon2016network}. There are many examples of FEPs, with varying security guarantees, including Shadowsocks \cite{shadow_socks_online}
and VMess-based \cite{vmess_online} systems.

Even though FEPs seem robust, they are easily recognizable when used without additional security measures, as "looking like nothing" is a standout feature. Recently, censors have been using a variety of methods ranging from statistical analysis to active probing to detect and block commercial FEPs such as Shadowsocks. Alice et al. \cite{alice2020china} demonstrated how Shadowsocks can be detected using the packet length and entropy. Later on, Wu et al. \cite{wu2023great} highlighted the extent of the detectability of FEPs, and were able to shed light on specific rules upon which FEPs are detected. They reveal that packets are blocked based on ad-hoc rules such as the number of printable ASCII characters, or the fact that the entropy of normal encrypted TLS packets is considerably lower than FEPs. These rules are subject to frequent changes, so the rigid design of older FEPs is unable to adapt to the updated detection rules. Additionally, Fenske et al. \cite{fenske2024bytes} deliver an extensive study on available FEPs and their features, highlighting their shortcomings, and revealing how packet lengths and protocol inconsistencies can be a main source of detection. They focus on fixing packet lengths using some primitive packet length shaping but leave the how to shape the traffic or shaping packet timings for future work. Although none of the aforementioned work discusses the effects of packet timings on the detection of FEPs, Wails et al. \cite{wails2024precisely} showcase the general importance of packet timings in detecting CC tools.

Fortunately, a solution for the challenges outlined above exists, a FEP-specific traffic shaper. Traffic shapers exist in other domains (e.g., IoT) to add a layer of privacy to the data in transit \cite{xiong2022network}. A FEP-specific traffic shaper could in theory shape data so it evades rule-based detection, helps normalize packet lengths, and control packet timings to thwart all previously mentioned attacks. Some prior work such as Proteus \cite{wails2023proteus} discusses how traffic shaping could help improve their tool's privacy guarantees, and some such as Fenske et al. \cite{fenske2024bytes} implement their own traffic shaper, but lack packet timing shaping and are not designed for integration with other solutions. Nevertheless, to the best of our knowledge, no prior stand-alone FEP traffic shaper exists that can address our concerns.

In this paper, we introduce Shaperd, a real-time traffic-shaping tool that can easily be integrated with any existing FEP. Shaperd provides a novel \textit{constraint} system, enabling its users to define the features of the generated traffic, shaping both the packets lengths and timings of a given traffic flow. Shaperd aims to be an independent traffic shaper that is compatible with all prior FEPs, thus addressing unobservability challenges highlighted by Wu et al. \cite{wu2023great}. The generality of Shaperd's constraint system enables users to generate traffic flows with any desired features, such as particular values for a packet's entropy, to circumvent detection systems. Moreover, this allows integration with previous FEPs detected using such methods while incurring minimal performance overheads.

Our main contributions are as follows: a) Shaperd, a traffic-shaping system that can easily be adopted by previous CC systems, allowing them to overcome detection techniques, and b) a novel constraint system that enables users to describe how they want to shape their traffic, resulting in a flexible traffic flow. 

\section{Technical Details}
\label{sec:techdetails}

In this section, we introduce Shaperd and its inner workings. Shaperd is designed to be easily integrated with any existing or future FEP, such as Shadowsocks \cite{ji2022security}. Shaperd relies on its constraint system to allow users to adapt to the ever-changing detection systems used by the adversaries, such as the ad-hoc rules used by the censor analyzed by Wu et al. \cite{wu2023great}. Shaperd uses two simple but effective methods to shape a packet's content, which was designed with constraint agnosticism in mind.

\mypara{Workflow. } The overall workflow of Shaperd is as follows. Alice, a Shaperd user, has a set of constraints outlining her desired traffic flow. This constraint set can either be created by her, received from trusted sources (e.g., how bridge information is distributed \cite{tulloch2023lox}), or a mix of both. Then, Alice starts a Shaperd instance with the constraint set and routes her FEP through Shaperd. Data received by Shaperd is stored in a traffic buffer prior to being shaped according to the defined constraints. As a part of this process, the input traffic goes through two separate systems before being sent out: the shaper and timer threads.

The shaper thread constantly monitors the traffic buffer, directly inspecting the contents of the buffered packets to generate new valid packets as soon as the contents either exceed the maximum capacity of a valid packet or a certain predefined period. Packets are created using two main shaping methods: a) length reduction, where a packet's content length gets reduced until the constraints are satisfied, and b) content padding, during which potential packets are padded with incremental byte values until a satisfying packet is found. We are using naive padding, leaving the exploration of more efficient heuristics for future versions (see a more detailed discussion in Appendix~\ref{app:additional-discussion}). Length reduction, although simple, has proven to be more effective and performative in our preliminary testing, so in our current version, content padding is used as a fallback when length reduction fails. Finally, the shaper sends all created packets to the timer thread. 


The timer thread is in charge of deciding when packets are sent out according to the preset timing constraints. These constraints can denote the flow throughput, the minimum and maximum delay between any two packets, or any other information regarding packet timing. We note that the time thread is currently a work in progress. Our goal is to support real-time shaping of packet timing by enforcing user-specified constraints on delay and throughput. We aim to finalize and test this feature in the near future. We discuss the timer and its state more thoroughly in \S\ref{sec:conclusion}.

\mypara{The constraint framework. } We present the constraint framework, a simple but effective system for defining content or timing-specific conditions on packets in traffic flows. A content constraint consists of four main components: a) the constraint function, a function that calculates the metric to be assessed, b) a value in the unit of the constraint function, which will be compared against, c) a mode, that sets the comparison operator (e.g., equal, lower than, or greater than), and d) a target, which denotes which packets this constraint should affect. We can support any constraint on a value expressible by a well-defined, deterministic function, however, constraints defined by non-computable functions, cannot be handled. A practical example of valid constraint could be one of the proposed rules described by Wu et al. \cite{wu2023great}, that there needs to be over 50\% of ASCII printable characters in a given packet. For this particular example, our constraint function would be one that calculates the percentage of ASCII printable characters in a given packet, the value would be 50\%, the mode would be equal to or greater than, and the target could be all packets. Timing constraints, on the other hand, are still a work in progress. We envision that users will be able to define constraints such as minimum and maximum delay between consecutive packets, fixed inter-packet intervals, or jitter bounds. These constraints will guide the timer thread in determining when to release packets from the queue, enabling realistic and tunable traffic shaping behavior.


\mypara{Implementation. } We have implemented a proof-of-concept of Shaperd~\cite{shaperd_online} and tested it using a mix of unit and real-time client-server tests. We developed our prototype using Go with $\sim$1000 lines of code and implemented a client that generates random bytes to act as the traffic generator in testing. In our tests, we mainly used a simple entropy constraint on messages to evaluate Shaperd's overhead. Our preliminary results reveal a minimal overhead of 4.1\% over the input throughput. This overhead scales with the number of additional constraints on the packets. Our evaluation revealed that the first constraint added an overhead of 5.1\% and the second added 5.5\% of overhead. We note that these values are subject to change, based on how rigorous the chosen constraints are.

\mypara{Usage scenarios. } We believe Shaperd can prove useful by not only helping widely deployed tools such as Shadowsocks and VMess-based systems achieve a new strong layer of privacy but also the development of new FEPs by shifting the concerns from traffic shaping to the development of other features and improvements.

\section{Discussion and Future Work}
\label{sec:conclusion}

As suggested by \S\ref{sec:techdetails}, Shaperd's performance has room for improvement. There are several features that we are actively working on. 

\mypara{Constraint agnosticism. } Shaperd's constraint agnostic design is both a strong point of our work and a major performance drawback. To accommodate this, we decided to keep our constraint agnostic design, while adding an optional field named "type" to our constraints. The type field can be used by the shaper to better understand the constraint and satisfy it more efficiently.  

\mypara{Supported protocols.} The current version of Shaperd only supports shaping TCP packets. We plan to add support for more protocols in the future to allow a wider array of tools to be at the disposal of our users and to allow more flexible traffic morphing strategies for evading modern detection techniques.

\mypara{Evaluation. } Our current evaluation methodology uses a random byte generator as the client. We aim to test Shaperd in combination with state-of-the-art academic tools such as Proteus \cite{wails2023proteus}, or widely adopted open-source tools such as Shadowsocks and V2Ray. Moreover, we plan to use sophisticated traffic fingerprinting techniques~\cite{shen2023subverting,wu2023great} to evaluate the effectiveness of our shaper.

To conclude, we introduce Shaperd, a CC-specific traffic shaper designed for seamless integration with existing FEPs. By leveraging a novel constraint system, our approach allows tool designers to offload the complexity of traffic shaping. While Shaperd is still a work in progress, we are committed to advancing its development and capabilities.

\bibliographystyle{ACM-Reference-Format}
\bibliography{sample-base}

\appendix

\section{Additional Discussion}
\label{app:additional-discussion}

\mypara{Padding algorithm.} We plan to use more prominent padding solutions, as most of the current performance loss stems from our naive padding system which tries to find proper padding bytes iteratively, which scales exponentially the more padding bytes a flow requires. This solution helps Shaperd's flexibility but is inefficient. We plan to use a heuristic-based padding system that selects padding bytes more efficiently in the future.

\mypara{Timing and blocking redirection.} To better support adaptive circumvention, we are exploring how Shaperd can detect and adapt to blocking. For detection, we can use tools such as Troll Patrol~\cite{trollpatrol_thesis}, which tackle various methods to detect endpoint blockages. Then, to adapt to blocking, Shaperd can adopt alternate constraint sets to alter packet patterns, thereby circumventing the block.

\end{document}